\documentclass[a4paper,twoside,10pt]{article}

\usepackage[english]{babel}
\usepackage[utf8x]{inputenc}
\usepackage[T1]{fontenc}

\usepackage[a4paper,top=1cm,bottom=2cm,left=1.5cm,right=1.5cm,marginparwidth=1.75cm]{geometry}

\usepackage{amsmath}
\usepackage{graphicx}
\usepackage[colorinlistoftodos]{todonotes}
\usepackage[colorlinks=true, allcolors=blue]{hyperref}
\usepackage{multicol}
\usepackage{authblk}
\usepackage[superscript,biblabel]{cite}
\usepackage{float}

\title{\large{\textbf{Tuning Rashba Spin--Orbit Coupling in Gated Multilayer
InSe}}}
\author[1]{\normalsize{Kasun Premasiri}}
\author[1]{Santosh Kumar Radha}
\author[1]{Sukrit Sucharitakul}
\author[2]{U. Rajesh Kumar}
\author[2,3]{Raman Sankar}
\author[2]{Fang-Cheng Chou}
\author[4,5]{Yit-Tsong Chen}
\author[1]{Xuan P. A. Gao\footnote{Corresponding author: xuan.gao@case.edu (X.P.A.G.)}}

\affil[1]{\textit{Department of Physics, Case Western Reserve University, 2076 Adelbert Road,
Cleveland, Ohio 44106, United States}} 
\affil[2]{\textit{Center for Condensed Matter Sciences, National Taiwan University, Taipei 10617, Taiwan}}
\affil[3]{\textit{Institute of Physics, Academia Sinica, Taipei 11529, Taiwan}}
\affil[4]{\textit{Department of Chemistry, National Taiwan University, Taipei 10617, Taiwan}} 
\affil[5]{\textit{Institute of Atomic and Molecular Sciences, Academia Sinica, Taipei 10617, Taiwan}}

\date{}

\begin{document}

\maketitle
\def\changemargin#1#2{\list{}{\rightmargin#2\leftmargin#1}\item[]}
\let\endchangemargin=\endlist 
\begin{changemargin}{1cm}{1cm} 
\noindent
Manipulating the electron spin with the aid of spin--orbit coupling (SOC) is an indispensable element of spintronics. Electrostatically gating a material with strong SOC results in an effective magnetic field which can in turn be used to govern the electron spin. In this work, we report the existence and electrostatic tunability of Rashba SOC in multilayer InSe. We observed a gate-voltage-tuned crossover from weak localization (WL) to weak antilocalization (WAL) effect in quantum transport studies of InSe, which suggests an increasing SOC strength. Quantitative analyses of magneto-transport studies and energy band diagram calculations provide strong evidence for the predominance of Rashba SOC in electrostatically--gated InSe. Furthermore, we attribute the tendency of the SOC strength to saturate at high gate voltages to the increased electronic density of states-induced saturation of the electric field experienced by the electrons in the InSe layer. This explanation of nonlinear gate voltage control of Rashba SOC can be generalized to other electrostatically--gated semiconductor nanomaterials in which a similar tendency of spin--orbit length saturation was observed (e.g., nanowire field effect transistors), and is thus of broad implications in spintronics. Identifying and controlling the Rashba SOC in InSe may serve pivotally in devising III--VI semiconductor-based spintronic devices in the future.
 
\end{changemargin} 

\begin{multicols}{2}
Two-dimensional (2D) materials have drawn much attention in the past decade because of their easily exfoliable nature; allowing researchers to isolate atomically--thin layers. This in turn, furnishes an ideal opportunity to obtain high-quality two-dimensional electron gases (2DEGs) in a variety of atomically--thin materials, beyond the conventional semiconductor-heterostructure-based scheme for the 2DEG formation\cite{art1,art2,art3,art4,art5,art6}. Thinning out certain 2D materials can unearth a host of fascinating phenomena owing to the quantum confinement, spin--orbit coupling (SOC) and changes in topology and symmetry of the crystal\cite{art7,art8,art9,art10,art11,art12}. Quantum transport studies work as a stepping stone to unravel many of those exotic phenomena\cite{art12,art13}. Furthermore, the ability to miniaturize 2D-material-based electronic devices while preserving quality, especially because of their layered van der Waals structures, has led to a lot of progress and excitement in 2D semiconductor nanoelectronics and optoelectronics\cite{art14,art15,art16,art17}.

\par
Indium monoselenide (InSe) is classified as one of the layered metal chalcogenides. The atomic structure of InSe is composed of individual layers that are interconnected through van der Waals forces having an interlayer distance ($d$) of about 0.8 nm (Figure 1a). Each layer has a honeycomb lattice with Se--In--In--Se atomic stacks that are covalently bonded. Being a monochalcogenide, InSe crystallizes in four different polytypes, namely $\epsilon$, $\gamma$, $\beta$, and $\delta$\cite{art18}. There are a few recent reports on exfoliated InSe. Carrier mobility of InSe has been reported to exceed 1000 cm\textsuperscript{2} V\textsuperscript{-1} s\textsuperscript{-1} at room temperature\cite{art19,art20} and 10\textsuperscript{4} cm\textsuperscript{2} V\textsuperscript{-1} s\textsuperscript{-1} at cryogenic temperatures\cite{art20}, asserting the importance of InSe in high-performance electronic devices. Moreover, quantum Hall effect has been observed in a high-quality 2DEG in few-layer InSe encapsulated with hexagonal boron nitride\cite{art20}. InSe becomes an important candidate not only from a charge-based devices perspective, but also for spin-based devices, because of its strong SOC. Therefore, InSe can potentially become a key candidate semiconductor in spintronics, where manipulating the electron spin works as the main functionality. In general, manipulating the electron spin using an electric field is preferred over its magnetic counterpart because of the ease in device operation. However, the SOC of InSe and its underlying mechanisms still remain as an unexplored territory due to the scarcity of experimental data on the subject. 

\par
SOC can be classified into two categories depending on its symmetry dependence. Symmetry-independent SOC exists in almost all types of crystal structures, because it stems from the spin--orbit interaction in atomic orbitals. However, symmetry-dependent SOC is not that commonplace, and exists only in crystal structures, which lack inversion symmetry. The latter of the above two categories involves two different types of SOC, namely Dresselhaus (due to bulk-induced asymmetry) and Rashba (due to surface- or interface-induced asymmetry) SOC\cite{art12}. The quintessence of the symmetry-dependent SOC is that electrons that are in motion in an electric field experience an effective magnetic field with respect to their frame of reference, even without any external magnetic fields present in the system. This pseudo magnetic field, which is also called the spin--orbit field gets coupled to the electron spin.  

\end{multicols}
\begin{figure}[ht]
\includegraphics[scale=0.087]{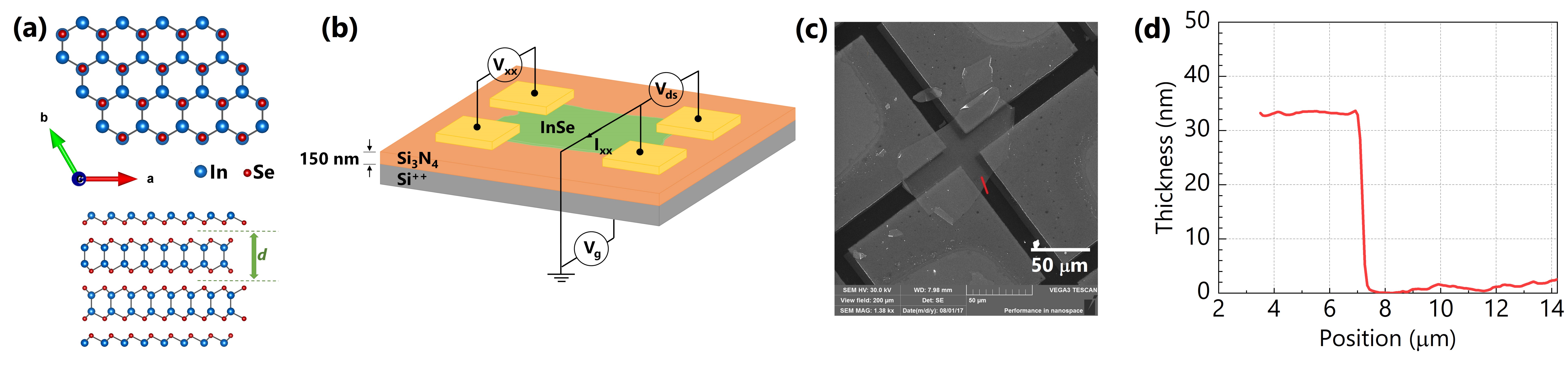}

\caption{\textbf{a.} Schematic of the atomic arrangement of InSe, \textbf{b.} Schematic of the device structure of the InSe field effect devices, \textbf{c.} Scanning electron micrograph of the InSe field effect device studied, \textbf{d.} AFM (atomic force microscope) thickness profile of the InSe device studied (along the red line marked in panel \textbf{c})}
\end{figure}

\begin{multicols}{2}

\noindent
Also the electric field that modifies the environment of the electrons could either arise intrinsically from the system per se or be externally applied. If this electric field is either externally applied or caused by the presence of a heterointerface, the resultant SOC becomes Rashba-type\cite{art21}. The Hamiltonian for the Rashba SOC is of the form:

\begin{equation} \label{eq:1}
{H}_{R} = \alpha_{R}(\boldsymbol{\sigma} \times \textbf{k}).\hat{E}
\end{equation}

\noindent
where, $\alpha_{R}$ is called the Rashba parameter; and $\boldsymbol{\sigma}$, \textbf{k}, and $\hat{E}$ are Pauli spin matrices, electron wave vector, and unit vector of the electric field, respectively. 

\par
Degenerately doped Si substrates with a layer of 150 nm thick Si\textsubscript{3}N\textsubscript{4} on the surface were used to fabricate InSe devices. To begin with, these substrates were sonicated in acetone for 20 min. Thereafter, they were cleaned with isopropyl alcohol followed by deionized water and were blow-dried with compressed air. Then, these cleaned substrates were treated in an ultraviolet-ozone cleaner at 150 °C for 10 min, to obtain a pristine Si\textsubscript{3}N\textsubscript{4} surface. InSe flakes were mechanically exfoliated from bulk $\beta$-type InSe on to cleaned Si\textsubscript{3}N\textsubscript{4} surfaces using scotch tape. Considering the thickness-dependent mobility of InSe\cite{art22}, a few exfoliated InSe flakes with different interference colors were profiled using an atomic force microscope (AFM), so that a correlation between the interference color and the corresponding thickness of the flake can be established to pick InSe flakes with thickness around 30 nm. Stencil lithography was used to fabricate metal contacts to exfoliated InSe flakes, with the aid of glider TEM grids (TED PELLA G200HS). Ti (15 nm) followed by Ag (50 nm) were deposited on TEM-grid-masked InSe samples using a metal evaporation system (Angstrom Evovac Deposition System).

\par
Figure 1b illustrates the device structure fabricated using the aforementioned specifics. The Si\textsubscript{3}N\textsubscript{4} layer between the InSe flake and the degenerately doped Si substrate serves as the gate dielectric for the purpose of electrostatic gating via a DC voltage applied to the Si substrate. van der Pauw geometry was used to perform transport and magneto-transport measurements (Figure 1b). Figure 1c is a scanning electron microscope (SEM) image of the device used for the magneto-transport analysis in this letter. According to the AFM profile in Figure 1d, the thickness of the sample is about 35 nm. Gold wires (50 $\mu$m in diameter) were soldered onto the metal contacts that are attached to the InSe flake (Figure 1b), and to the sample holder of the physical property measurement system (PP-

\begin{figure}[H]
\centering
\includegraphics[scale=0.095]{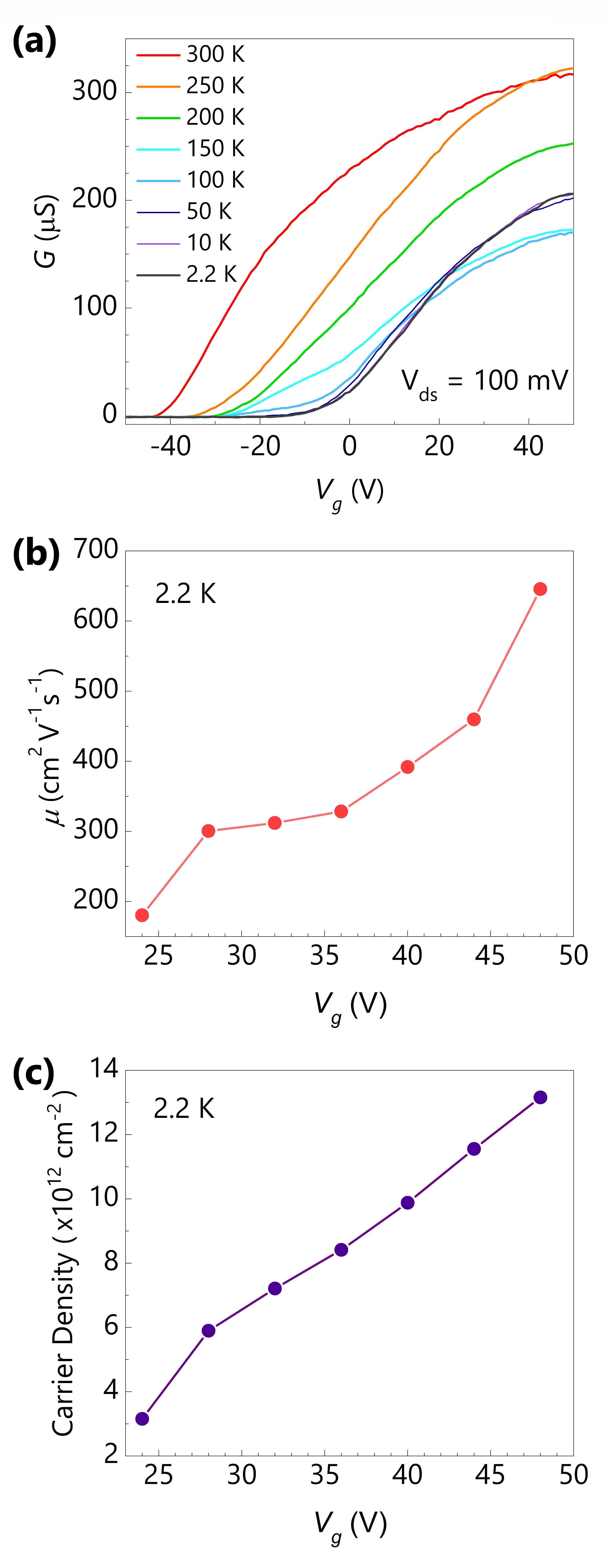}

\caption{\textbf{a.} Two-probe conductance vs gate voltage at different temperatures, \textbf{b.} Electron mobility (Hall mobility) of the device as a function of the gate voltage at 2.2 K, \textbf{c.} Carrier density of the device as a function of the gate voltage at 2.2 K}
\end{figure}

\end{multicols}
\begin{figure}[ht]
\centering
\includegraphics[scale=0.0969]{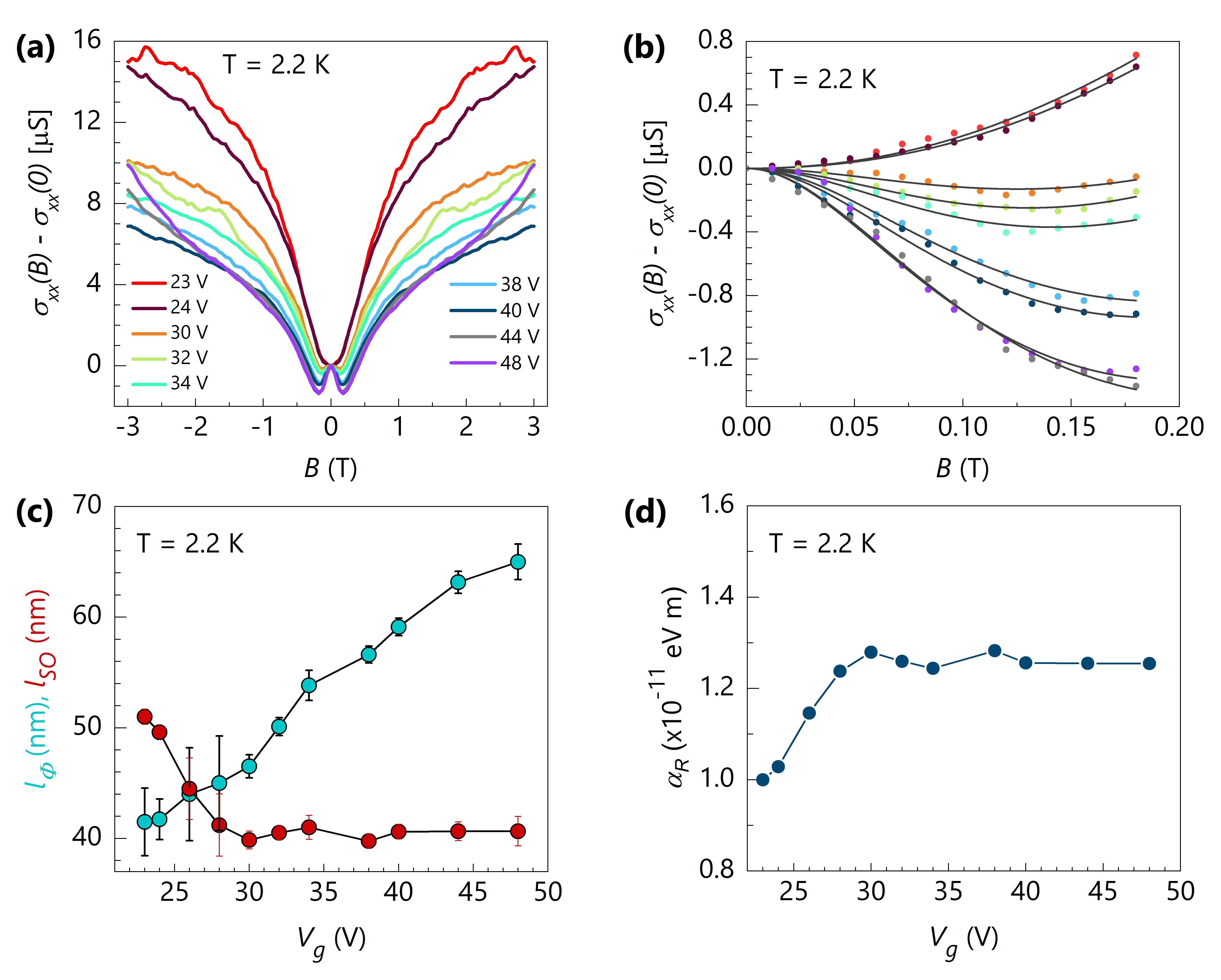}

\caption{\textbf{a.} Magneto-conductance of the InSe field effect device under different gate voltages when a perpendicular magnetic field is applied (at 2.2 K), \textbf{b.} Fitting for the magneto-conductance response in \textbf{a} using the HLN equation, \textbf{c.} Phase coherence length and spin--orbit relaxation length as a function of the applied gate voltage at 2.2 K, \textbf{d.} Rashba SOC strength as a function of the applied gate voltage at 2.2 K}
\end{figure}

\begin{multicols}{2}

\noindent
MS) [Quantum Design, Model 6000]. All the transport measurements were performed in the PPMS using either DC or lock-in technique. 

\par
The response of the two-probe conductance (\textit{G}) to the back-gate voltage (\textit{V\textsubscript{g}}) at different temperatures was explored at the onset to acquire a basic overview of the sample (Figure 2a). The conductance was measured at a DC source-drain bias of \textit{V\textsubscript{ds}} = 0.1 V. Although the two-probe measurement of conductance does not reflect the intrinsic conductance due to the contact resistance, the basic gate transfer curves in Figure 2a confirm that the InSe device is comprised of well-behaved, n-type field-effect transistor (FET) characteristics, with gate-tunable electrical properties. Using four-probe Hall measurements, electron Hall mobility and electron density as a function of the gate voltage applied to the InSe device (Figure 1b) were measured (Figure 2b,c). It can be clearly seen that both the Hall mobility and carrier density increase with the gate voltage. For the range of gate voltages used in the magneto-transport measurements, InSe exhibited good Ohmic contacts, and the corresponding carrier density covers a broad range, 10\textsuperscript{12} -- 10\textsuperscript{13} cm\textsuperscript{-2}.  

\par
Magneto-conductance measurements at low temperatures (from 50 to 2.2 K) were employed to investigate the SOC effect and the other pertinent parameters for InSe nanoflakes. All the magneto-conductance measurements were symmetrized for $\sigma_{xx}$ and antisymmetrized for $\sigma_{xy}$ to get rid of any possible mixing between $\sigma_{xx}$ and $\sigma_{xy}$ due to the imperfect alignment of contacts. With increasing \textit{V\textsubscript{g}}, InSe exhibits a transition from weak localization (WL) to weak antilocalization (WAL) at about \textit{V\textsubscript{g}} = 30 V (Figure 3a). This crossover has been observed by tuning either the gate voltage (Figure 3a) or the temperature of the system (Figure 4a). Quantum transport of electronic systems in diffusive regime can be used to elucidate the WAL and WL. Coherent backscattering (constructive interference) between identical but time-reversed diffusive paths of electrons in a pool of randomly distributed scatters, leads to a negative correction to the conductance, and this phenomenon is called WL\cite{art23}. However, strong SOC causes those identical, time-reversed diffusive loops to acquire a phase change of 2$\pi$, because of the coupling between the effective magnetic field and the electron spin. This suppresses coherent backscattering and leads to WAL. In this study, we used the Hikami--Larkin--Nagaoka (HLN) equation to extract important parameters in the quantum transport process of electrons\cite{art24}. The magneto-conductivity is

\begin{equation} \label{eq:2}
\begin{split}
\sigma(B)-\sigma(0) = \frac{e^2}{2\pi^2\hbar}[\textit{ln}(\frac{B_{\phi}}{B})-\Psi(\frac{1}{2}+\frac{B_{\phi}}{B})]\\+\frac{e^2}{\pi^2\hbar}[\textit{ln}(\frac{B_{SO}+B_{e}}{B})-\Psi(\frac{1}{2}+\frac{B_{SO}+B_{e}}{B})]\\-\frac{3e^2}{2\pi^2\hbar}[\textit{ln}(\frac{(4/3)B_{SO}+B_{\phi}}{B})-\Psi(\frac{1}{2}+\frac{(4/3)B_{SO}+B_{\phi}}{B})]
\end{split}
\end{equation}

\noindent
where, $\Psi(x)$ is the digamma function and $B_{i}=\hbar/4el_{i}^2$; $l_{\phi}$ 

\end{multicols}
\begin{figure}[ht]
\centering
\includegraphics[scale=0.1]{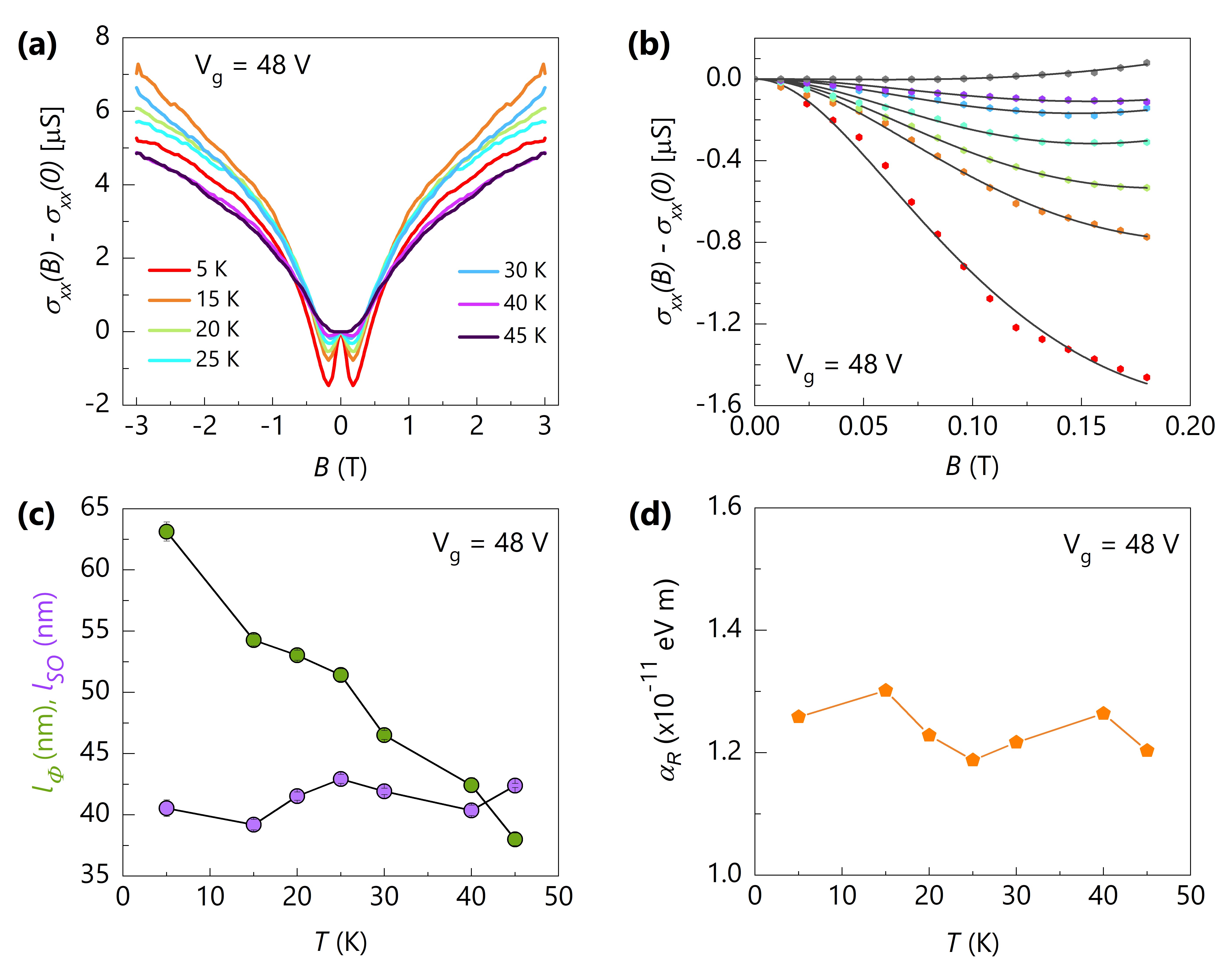}

\caption{\textbf{a.} Magneto-conductance of the InSe field effect device under different temperatures when a perpendicular magnetic field is applied (at \textit{V\textsubscript{g}} = 48 V), \textbf{b.} Fitting for the magneto-conductance response in \textbf{a} using the HLN equation, \textbf{c.} Phase coherence length and spin--orbit relaxation length as a function of temperature at \textit{V\textsubscript{g}} = 48 V, \textbf{d.} Rashba SOC strength as a function of temperature at \textit{V\textsubscript{g}} = 48 V}
\end{figure}

\begin{multicols}{2}

\noindent
is the phase coherence length, $l_{SO}$ is the spin--orbit relaxation length, and $l_{e}$ is the mean free path. Figure 3b and Figure 4b show the fitting performed using the HLN equation. Note that the fitting was limited to the cases where the conductivity of the sample was above $2e^2/h$ (=77.48 $\mu$S) for the sample to be in the diffusive regime. The range for the magnetic field used for the fitting was confined to be less than about 200 mT to maintain the validity of the equation with respect to the characteristic magnetic fields on it. Additionally, the fitting appears to be in good agreement with the experimental data. At higher magnetic fields beyond the WL or WAL regime, there is a positive magneto-conductance or negative magneto-resistance, which might be related to Coulomb interaction effect induced negative magneto-resistance.

\par
The HLN equation was used to extract two parameters, namely, the phase coherence length ($l_{\phi}$) and the spin--orbit relaxation length ($l_{SO}$). According to Figure 3c, $l_{\phi}$ increases with the gate voltage from about 42 to 62 nm. This is a direct consequence of the increased carrier density, and thus Fermi velocity and diffusion constant, with electrostatic gating. Additionally, with temperature, $l_{\phi}$ changes from about 63 to 38 nm; as there is more inelastic (electron--electron, electron--phonon) scattering, causing dephasing at higher temperatures  (Figure 4c). As for $l_{SO}$, it is more or less constant with temperature (Figure 4c). However, with the gate voltage, it shows a decreasing trend first, followed by a saturation behavior (Figure 3c). In the remainder of this Letter, the focus is on discussing and understanding the behavior of $l_{SO}$ in relation to the Rashba effect. In the literature, it is customary to use $l_{SO}$ to calculate the Rashba parameter ($\alpha_{R}$)\cite{art25}:

\begin{equation} \label{eq:3}
\alpha_{R} = \frac{\hbar^2}{2ml_{SO}}
\end{equation}

\noindent
Figure 3d and Figure 4d illustrate the variation of the inferred Rashba parameter with the gate voltage and temperature, respectively. While the weak temperature dependence of $\alpha_{R}$ in Figure 4d seems reasonable, the saturation behavior of $\alpha_{R}$ at higher gate voltages in Figure 3d is puzzling. Because, according to Equation 1, $\alpha_{R}$ is expected to be proportional to the electrical field.

\par
Energy band diagram calculations were performed to delve into the saturation behavior of the spin--orbit relaxation length ($l_{SO}$) -- and thus the Rashba parameter ($\alpha_{R}$) -- with the gate voltage. Poisson’s and Schrödinger equations were solved self-consistently, in order to obtain the energy band variations across the interface between InSe and Si\textsubscript{3}N\textsubscript{4}; as well as the electron wave function perpendicular to the 2D plane, to calculate the mean electrical field experienced by the electrons in InSe at different gate voltages. Material properties of InSe such as band offsets with Si\textsubscript{3}N\textsubscript{4} and doped Si, dielectric constant, bandgap, and band-edge effective masses were used for the calculation. Figure 5a illustrates how the conduction band energy of InSe (along the \textit{z}-direction) responds to electrostatic g-

\par

\end{multicols}
\begin{figure}[ht]
\centering
\includegraphics[scale=0.15]{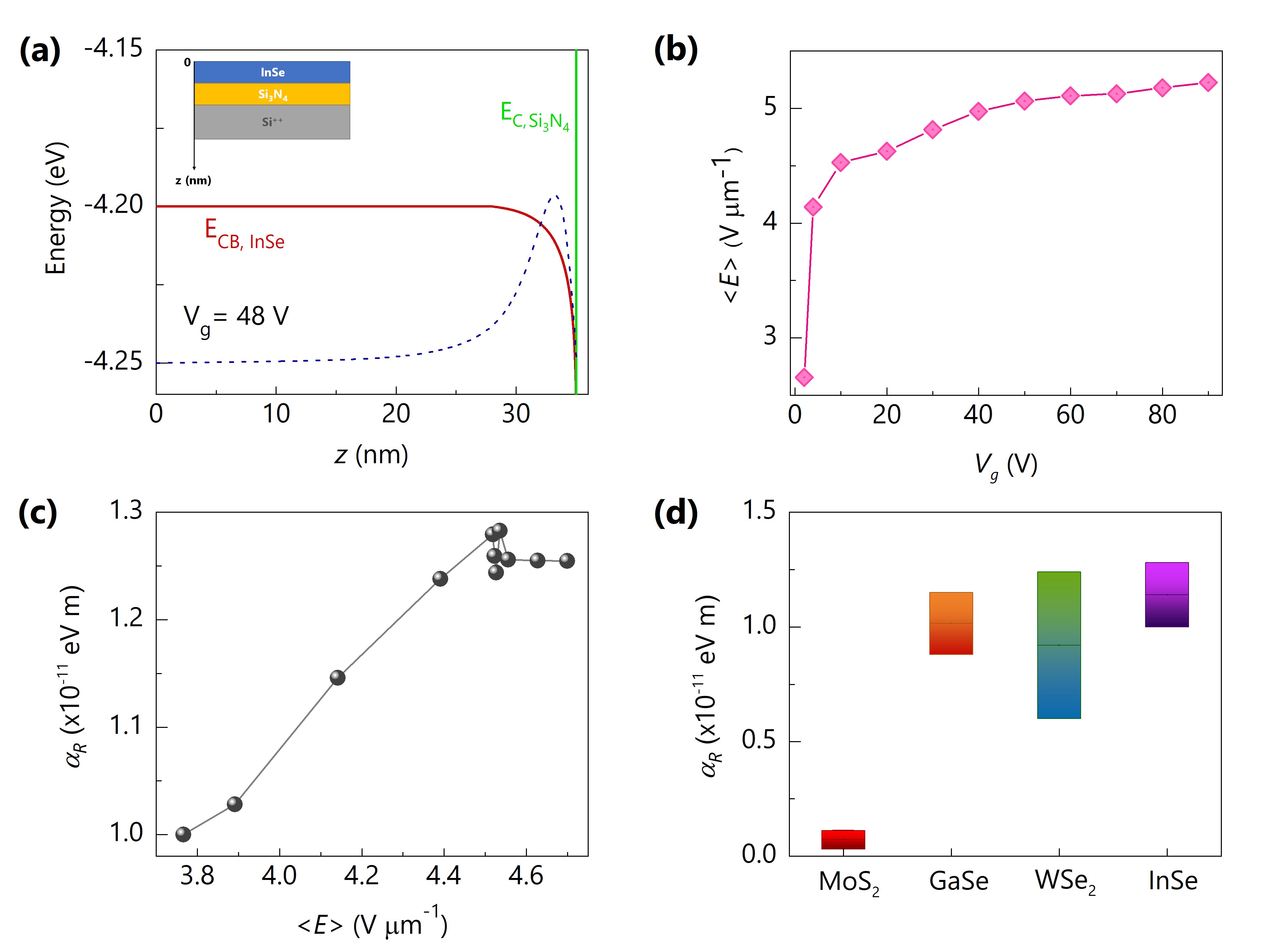}

\caption{\textbf{a.} The energy band diagram for InSe at \textit{V\textsubscript{g}} = 48 V (dashed line: spatial variation of the total density electron wave function, inset: schematic side view of the structure defining the \textit{z} direction), \textbf{b.} Average electric field (\textit{<E>}) across the InSe layer vs the applied gate voltage (\textit{V\textsubscript{g}}), \textbf{c.} Rashba parameter ($\alpha_{R}$) vs \textit{<E>}, \textbf{d.} A comparison of the Rashba parameter ($\alpha_{R}$) of MoS\textsubscript{2}\cite{art29}, GaSe\cite{art30}, and WSe\textsubscript{2}\cite{art13} reported in the literature, with  InSe in this work}
\end{figure}

\begin{multicols}{2}

\noindent
ating at \textit{V\textsubscript{g}} = 48 V. The way the energy band diagrams respond to changing the gate voltage bias indicates InSe lacks mobile carriers initially. This is due to the difference in work functions between InSe and doped Si. When an adequately large positive bias is applied to the Si gate layer, the conduction band of InSe exhibits a downward bending and a 2DEG forms in the InSe layers near the InSe and Si\textsubscript{3}N\textsubscript{4} interface.

\par
In this regime, electrostatically gating the FET structure of InSe results in a potential drop across the InSe nanoflake, causing a finite average electric field to form out of the 2D plane. This is a direct consequence of the finite quantum capacitance of the 2DEG at the InSe/ Si\textsubscript{3}N\textsubscript{4} interface, and of the imperfect screening of the gate electric field by the 2DEG. Thus the applied electric field only undergoes partial screening from InSe. Figure 5b illustrates the response of the average electric field: \textit{<E>} (which was calculated using the potential drop weighted by the electron wave function distribution across the InSe nanoflake) to the applied gate voltage. This response reveals that the average electric field tends to saturate with increasing gate voltage, reflecting the better screening ability of 2DEGs at high electron densities. On the other hand, this average electric field functions as the driving force for the tuning of SOC in InSe. Therefore, it appears that the standard ways of electrostatic gating in FETs (for instance, back-gating or top-gating), do not provide indefinite tuning of SOC, rather it becomes limited in its capacity at higher voltages. A similar kind of saturation behavior in spin--orbit relaxation length has been observed for other nanomaterials (e.g., semiconductor nanowires) as well, when the gate voltage simultaneously modulates both the carrier density and the electrical field at the semiconductor/ dielectric interface for the Rashba SOC\cite{art25,art26,art27,art28}. We believe that the essential physics of those systems is the same as in this study. Hence, to avoid the apparent nonlinear gate voltage dependence of SOC strength and saturation of Rashba SOC strength, a gating scheme that can separate the control of interface electric field from the carrier density modulation is needed.  For instance, sandwiching the semiconductor with two floating gates would allow the perpendicular electric field across the semiconductor to be controlled without changing the carrier density in the semiconductor. Indeed, such double-gate configuration was employed to generate and tune the Rashba SOC in InAs nanowires, and $\alpha_{R}$ was seen to vary linearly with the increasing voltage between top and back gates without any sign of saturation\cite{art25}.

\par
The variation of the Rashba parameter ($\alpha_{R}$) with the average electric field (<E>) is plotted in Figure 5c. It shows that the Rashba parameter increases proportionally to the average electric field, throughout the gate voltage range used. This corroborates that the saturation of the average electric field inside the InSe nanoflake (in the FET structure) is responsible for the saturation behavior of the spin--orbit relaxation length against the gate voltage in Figure 3c. Hence, the SOC of InSe is dominated by Rashba SOC. Figure 5d illustrates where InSe stands compared to some other 2D semiconductors (MoS\textsubscript{2}\cite{art29}, GaSe\cite{art30}, and WSe\textsubscript{2}\cite{art13}), in terms of the Rashba parameter (values obtained in this work together with the values reported in literature). This confirms that the strength of $\alpha_{R}$ in InSe is comparable to the highest $\alpha_{R}$ reported in some other 2D chalcogenides.

\par
To conclude, we have observed gate-controlled WAL and WL behaviors in multilayer InSe. The extracted values for the phase coherence length and spin--orbit relaxation length can be used to understand the spin relaxation and electron decoherence in InSe. Energy band diagram calculations reveal that the Rashba SOC dominates the SOC in back-gated InSe. Being a 2D material with high electron mobility, combined with strong SOC, InSe appears to be a compelling candidate for future spintronic applications. 

\vspace{15mm}

\noindent
X.P.A.G. thanks the National Science Foundation for its financial support under Award DMR-1151534. Y.T.C. acknowledges the financial support from the Ministry of Science and Technology of Taiwan under MOST 106-2627-M-002-035 and MOST 105-2119-M-032-002.
%
\vspace{10mm}

\noindent
\textbf{\uppercase{References}}
\begingroup
\renewcommand{\section}[2]{}%

\bibliographystyle{unsrt}
\bibliography{references}

\begin{thebibliography}{10}

\bibitem{art1}
Li~L. et~al.
\newblock Quantum \uppercase{H}all effect in black phosphorus two-dimensional
  electron system.
\newblock {\em Nature Nanotech. \textbf{11}}, 593-597 (2016).

\bibitem{art2}
Gillgren N. et~al.
\newblock Gate tunable quantum oscillations in air-stable and high mobility
  few-layer phosphorene heterostructures.
\newblock {\em 2D Mater. \textbf{2}}, 011001 (2015).

\bibitem{art3}
Li~L. et~al.
\newblock Quantum oscillations in a two-dimensional electron gas in black
  phosphorus thin films.
\newblock {\em Nature Nanotech. \textbf{10}}, 608-613 (2015).

\bibitem{art4}
Fallahazad B. et~al.
\newblock \uppercase{S}hubnikov–de \uppercase{H}aas \uppercase{o}scillations
  of \uppercase{h}igh-\uppercase{m}obility \uppercase{h}oles in
  \uppercase{m}onolayer and \uppercase{b}ilayer
  \uppercase{W}\uppercase{S}e\textsubscript{2}: \uppercase{L}andau
  \uppercase{l}evel \uppercase{d}egeneracy, \uppercase{e}ffective
  \uppercase{m}ass, and \uppercase{n}egative \uppercase{c}ompressibility.
\newblock {\em Phys. Rev. Lett. \textbf{116}}, 086601 (2016).

\bibitem{art5}
Xu~S. et~al.
\newblock Universal low-temperature \uppercase{o}hmic contacts for quantum
  transport in transition metal dichalcogenides.
\newblock {\em 2D Mater. \textbf{3}}, 021007 (2016).

\bibitem{art6}
Cui X. et~al.
\newblock Multi-terminal transport measurements of
  \uppercase{M}o\uppercase{S}\textsubscript{2} using a van der
  \uppercase{W}aals heterostructure device platform.
\newblock {\em Nature Nanotech. \textbf{10}}, 534-540 (2015).

\bibitem{art7}
Splendiani A. et~al.
\newblock Emerging \uppercase{P}hotoluminescence in \uppercase{m}onolayer
  \uppercase{M}o\uppercase{S}\textsubscript{2}.
\newblock {\em Nano Lett. \textbf{10}}, 1271-1275 (2010).

\bibitem{art8}
Castro~Neto A.H., Guinea F., Peres N.M.R., Novoselov K.S., and Geim A.K.
\newblock The electronic properties of graphene.
\newblock {\em Rev. Mod. Phys. \textbf{81}}, 109-162 (2009).

\bibitem{art9}
Kim Y. et~al.
\newblock Anomalous \uppercase{R}aman scattering and lattice dynamics in mono-
  and few-layer \uppercase{W}\uppercase{T}e\textsubscript{2}.
\newblock {\em Nanoscale \textbf{8}}, 2309-2316 (2016).

\bibitem{art10}
Fei Z. et~al.
\newblock Edge conduction in monolayer
  \uppercase{W}\uppercase{T}e\textsubscript{2}.
\newblock {\em Nature Phys. \textbf{13}}, 677-682 (2017).

\bibitem{art11}
Tang S. et~al.
\newblock Quantum spin \uppercase{H}all state in monolayer
  1\uppercase{T}'-\uppercase{W}\uppercase{T}e\textsubscript{2}.
\newblock {\em Nature Phys. \textbf{13}}, 683-687 (2017).

\bibitem{art12}
Manchon A. et~al.
\newblock New perspectives for \uppercase{R}ashba spin–orbit coupling.
\newblock {\em Nature Mater. \textbf{14}}, 871-882 (2015).

\bibitem{art13}
Yuan H. et~al.
\newblock Zeeman-type spin splitting controlled by an electric field.
\newblock {\em Nature Phys. \textbf{9}}, 563-569 (2013).

\bibitem{art14}
Jariwala D. et~al.
\newblock Emerging \uppercase{d}evice \uppercase{a}pplications for
  \uppercase{s}emiconducting \uppercase{t}wo-\uppercase{d}imensional
  \uppercase{t}ransition \uppercase{m}etal \uppercase{d}ichalcogenides.
\newblock {\em ACS Nano \textbf{8}}, 1102–1120 (2014).

\bibitem{art15}
Radisavljevic B. et~al.
\newblock Single-layer \uppercase{M}o\uppercase{S}\textsubscript{2}
  transistors.
\newblock {\em Nature Nanotech. \textbf{6}}, 147–150 (2011).

\bibitem{art16}
Wang Q.H. et~al.
\newblock Electronics and opto-electronics of two-dimensional transition metal
  dichalcogenides.
\newblock {\em Nature Nanotech. \textbf{7}}, 699–712 (2012).

\bibitem{art17}
Fiori G. et~al.
\newblock Electronics based on two-dimensional materials.
\newblock {\em Nature Nanotech. \textbf{9}}, 768–779 (2014).

\bibitem{art18}
Do~D.T. et~al.
\newblock Spin splitting in 2\uppercase{D} monochalcogenide semiconductors.
\newblock {\em Sci. Rep. \textbf{5}}, 17044 (2015).

\bibitem{art19}
Sucharitakul S. et~al.
\newblock \uppercase{I}ntrinsic \uppercase{e}lectron \uppercase{m}obility
  \uppercase{e}xceeding 10\textsuperscript{3}
  cm\textsuperscript{2}/(\uppercase{V}s) in \uppercase{m}ultilayer
  \uppercase{I}n\uppercase{S}e \uppercase{FET}s.
\newblock {\em Nano Lett. \textbf{15}}, 3815–3819 (2015).

\bibitem{art20}
Bandurin D.A. et~al.
\newblock High electron mobility, quantum \uppercase{H}all effect and anomalous
  optical response in atomically thin \uppercase{I}n\uppercase{S}e.
\newblock {\em Nature Nanotech. \textbf{12}}, 223–227 (2017).

\bibitem{art21}
Bychkov Y.A. and Rashba E.I.
\newblock Properties of a 2\uppercase{D} electron gas with lifted spectral
  degeneracy.
\newblock {\em P. Zh. Eksp. Teor. Fiz. \textbf{39}}, 66-69 (1984).

\bibitem{art22}
Feng W. et~al.
\newblock Performance improvement of multilayer \uppercase{I}n\uppercase{S}e
  transistors with optimized metal contacts.
\newblock {\em Phys. Chem. Chem. Phys. \textbf{17}}, 3653-3658 (2015).

\bibitem{art23}
Chakravarty S. and Schmid A.
\newblock Weak localization: The quasiclassical theory of electrons in a random
  potential.
\newblock {\em Phys. Rep. \textbf{140}}, 193-236 (1986).

\bibitem{art24}
Hikami S., Larkin A.I., and Nagaoka Y.
\newblock Spin-\uppercase{O}rbit \uppercase{I}nteraction and
  \uppercase{M}agnetoresistance in the \uppercase{T}wo \uppercase{D}imensional
  \uppercase{R}andom \uppercase{S}ystem.
\newblock {\em Prog. Theor. Phys. \textbf{63}}, 707-710 (1980).

\bibitem{art25}
Liang D. and Gao X.P.A.
\newblock Strong \uppercase{t}uning of \uppercase{R}ashba
  \uppercase{S}pin–\uppercase{O}rbit \uppercase{I}nteraction in
  \uppercase{S}ingle \uppercase{I}n\uppercase{A}s \uppercase{N}anowires.
\newblock {\em Nano Lett. \textbf{12}}, 3263–3267 (2012).

\bibitem{art29}
Neal A.~T. et~al.
\newblock Magneto-transport in \uppercase{M}o\uppercase{S}\textsubscript{2}:
  \uppercase{P}hase \uppercase{C}oherence, \uppercase{S}pin–\uppercase{O}rbit
  \uppercase{S}cattering, and the \uppercase{H}all \uppercase{F}actor.
\newblock {\em ACS Nano \textbf{7}}, 7077–7082 (2013).

\bibitem{art30}
Takasuna S. et~al.
\newblock Weak antilocalization induced by \uppercase{R}ashba spin-orbit
  interaction in layered \uppercase{III}-\uppercase{VI} compound semiconductor
  \uppercase{G}a\uppercase{S}e thin films.
\newblock {\em Phys. Rev. B \textbf{96}}, 161303 (2017).

\bibitem{art26}
Hansen A.E. et~al.
\newblock \uppercase{S}pin relaxation in \uppercase{I}n\uppercase{A}s nanowires
  studied by tunable weak antilocalization.
\newblock {\em Phys. Rev. B \textbf{71}}, 205328 (2005).

\bibitem{art27}
Hao X-.J. et~al.
\newblock \uppercase{S}trong and \uppercase{T}unable
  \uppercase{S}pin-\uppercase{O}rbit \uppercase{C}oupling of
  \uppercase{O}ne-\uppercase{D}imensional \uppercase{H}oles in
  \uppercase{G}e/\uppercase{S}i \uppercase{C}ore/\uppercase{S}hell
  \uppercase{N}anowires.
\newblock {\em Nano Lett. \textbf{10}}, 2956–2960 (2010).

\bibitem{art28}
Takase K. et~al.
\newblock \uppercase{H}ighly gate-tuneable \uppercase{R}ashba spin-orbit
  interaction in a gate-all-around \uppercase{I}n\uppercase{A}s nanowire
  metaloxide-semiconductor field-effect transistor.
\newblock {\em Sci. Rep. \textbf{7}}, 930 (2017).

\end{thebibliography}

\endgroup
\end{multicols}

\end{document}